\documentstyle[seceq,mbf,preprint]{ptptex}
\notypesetlogo 
\title{%
Lorentz-Invariant Non-Commutative Space-Time\\
Based On DFR Algebra
}
\vspace{5mm}
\author{%
  Hiromi {\sc Kase}$^{1}$, Katsusada {\sc Morita}$^{2}$,\\
  Yoshitaka {\sc Okumura}$^{3}$ and Eizou Umezawa$^{4}$
}
\inst{%
{\it  $^{1}$Department of Physics, 
  Daido Institute of Technology, Nagoya 457-0811, Japan\\
  $^{2}$Department of Physics, 
  Nagoya University, Nagoya 464-8602, Japan\\
  $^{3}$Department of Natural Sciences, 
  Chubu University, Kasugai, 487-0027, Japan\\
  $^{4}$School of Health Sciences, Fujita Health University, 
  Toyoake 470-1192, Japan}
}
\vspace{5mm}
\abst{%
It is argued that the familiar algebra of the non-commutative 
space-time with $c$-number $\theta^{\mu\nu}$ is inconsistent
from a theoretical point of view.
Consistent algebras are obtained by promoting
$\theta^{\mu\nu}$ to an anti-symmetric tensor operator 
${\hat\theta}^{\mu\nu}$.
The simplest among them
is 
Doplicher-Fredenhagen-Roberts 
(DFR) algebra in which 
the triple commutator
among the coordinate operators
is assumed to vanish.
This
allows us to define the Lorentz-covariant operator 
fields on the DFR algebra
as operators diagonal in the 
6-dimensional $\theta$-space of the hermitian
operators, ${\hat\theta}^{\mu\nu}$.
It is shown that
we then recover Carlson-Carone-Zobin (CCZ)
formulation of the Lorentz-invariant
non-commutative gauge theory
with no need of compactification
of the extra 6 dimensions.
It is also pointed out that
a general argument concerning 
the normalizability
of the weight function
in the Lorentz metric 
leads to
a division of
the $\theta$-space into two
disjoint spaces not connected by any Lorentz
transformation
so that
the CCZ covariant moment formula
holds true in each space, separately.
A non-commutative generalization
of Connes' two-sheeted Minkowski
space-time is also proposed.
Two simple models of quantum field theory
are reformulated on $M_4\times Z_2$
obtained in the commutative limit.
}
\begin{document}
\maketitle
\section{Introduction}
The Lorentz invariance is one of the fundamental symmetries
in (relativistic) quantum field theory (QFT).
Even if the Minkowski space-time would become
no longer a continuum but could be described by
a non-commutative geometry at, say, Planck scales,
the Lorentz invariance
should be maintained simply because
only Lorentz-covariant fields
are qualified to represent
interactions between matters and fields.
There is a long history\cite{1)} about the existence
of a minimum length in relation with quantum mechanics, 
special relativity and general relativity.
Such a minimum length, if exists,
should be introduced into the theory
without conflict to relativistic invariance.
\\
\indent
From this point of view the popular non-commutative 
space-time
characterized by the algebra,
\begin{eqnarray}
[{\hat x}^\mu,{\hat x}^\nu]&=&i\theta^{\mu\nu},
\label{eqn:1-1}
\end{eqnarray}
where ${\hat x}^\mu$
is the coordinate operators
and $\theta^{\mu\nu}$ is a real, anti-symmetric constant,
is unacceptable as a possible space-time
structure at short distances
because it violates the Lorentz
symmetry.
The apparent lack of the Lorentz
covariance of the algebra is cured by considering
a {\it set} of
similar algebras
as if $\theta^{\mu\nu}$
is a 2-tensor.
Namely, if one assumes
the algebra (\ref{eqn:1-1})
in a particular
Lorentz frame,
the algebra in a general
Lorentz frame connected by
a Lorentz transformation $(\Lambda^\mu_{\;\nu})$
with the original one
could be given
by
\begin{eqnarray}
[{\hat x}'{}^\mu,{\hat x}'{}^\nu]&=&i\theta'{}^{\mu\nu},
\label{eqn:1-2}
\end{eqnarray}
where ${\hat x}'=
\Lambda^{-1}{\hat x}$
and $\theta'=
(\Lambda^{-1})^2
\theta$.
If interpreted this way,
we call (\ref{eqn:1-1}) the
$c$-number $\theta$-algebra.
As we shall see in the
next section,
this interpretation
of the algebra (\ref{eqn:1-1})
is still not consistent theoretically.
It is clear, however, that,
in any case,
(\ref{eqn:1-1})
automatically implies the vanishing
triple commutator,
\begin{eqnarray}
[{\hat x}^\mu,[{\hat x}^\nu,{\hat x}^\rho]]&=&0,
\label{eqn:1-3}
\end{eqnarray}
which becomes a nontrivial constraint
in a non-commutative space-time
in which the commutator 
$[{\hat x}^\mu,{\hat x}^\nu]$
is no longer assumed to be a $c$-number.
\\
\indent
In these respects it should be recalled that
Snyder\cite{2)} was the first to
realize that the Lorentz invariance
does not necessarily require a continuum space-time.
Instead he invented a non-commutative space-time
with a fundamental length,
where the coordinates are no longer commutative
but obey the following commutation 
relations,
\begin{eqnarray}
[{\hat x}^\mu,{\hat x}^\nu]&=&ia^2{\hat M}^{\mu\nu},\nonumber\\[2mm]
[{\hat M}^{\mu\nu},{\hat x}^\rho]&=&i(g^{\nu\rho}{\hat x}^\mu
-g^{\mu\rho}{\hat x}^\nu),\nonumber\\[2mm]
[{\hat M}^{\mu\nu},{\hat M}^{\rho\sigma}]&=&
i(g^{\nu\rho}{\hat M}^{\mu\sigma}
-g^{\nu\sigma}{\hat M}^{\mu\rho}-g^{\mu\rho}{\hat M}^{\nu\sigma}
+g^{\mu\sigma}{\hat M}^{\nu\rho}).
\label{eqn:1-4}
\end{eqnarray}
Here, $a$ is a fundamental length
in the theory and ${\hat M}^{\mu\nu}$
are the infinitesimal generators
of the Lorentz group,
the metric being given by
$(g^{\mu\nu})=(+1,-1,-1,-1)$.
It is important to recognize
that the commutator
$[{\hat x}^\mu,{\hat x}^\nu]$
is a $q$-number unlike (\ref{eqn:1-1}).
A short look at (\ref{eqn:1-4})
shows that Snyder's quantized space-time\footnote{Snyder's 
motivation to introduce a quantized 
space-time is to
modify point interactions between fields and matters
so that UV divergence trouble in relativistic QFT
may be avoided. However, a storm of applause in the 
success of the renormalization theory 
puts aside Snyder's important paper which was largely ignored
in physics community.}
is Lorentz-invariant.
The relativistic QFT on it was developed in Ref.~3)
in which a simple geometric interpretation
of Snyder's quantized space-time in terms
of curved momentum space of constant curvature
was also described.
We here note that, in contrast to (\ref{eqn:1-3}),
the triple commutator in Snyder's quantized 
space-time is no longer vanishing,
\begin{eqnarray}
[{\hat x}^\mu,[{\hat x}^\nu,{\hat x}^\rho]]
=-a^2(g^{\nu\mu}{\hat x}^\rho
-g^{\rho\mu}{\hat x}^\nu).
\label{eqn:1-5}
\end{eqnarray}
Such a $q$-number triple commutator
is not a general feature
of a Lorentz-invariant non-commutative
space-time.
\\
\indent
In fact,
Doplicher, Fredenhagen and Roberts (DFR)
were led\cite{4)} to propose a new algebra
of a non-commutative space-time
through considerations on the space-time uncertainty relations
derived from quantum mechanics and
general relativity.
In the DFR algebra $\theta^{\mu\nu}$
in (\ref{eqn:1-1})
is replaced with an 
anti-symmetric tensor operator ${\hat\theta}^{\mu\nu}$
which simply {\it defines} the commutator
$-i[{\hat x}^\mu,{\hat x}^\nu]$.
DFR further assumed the vanishing
of the triple commutator
as in (\ref{eqn:1-3}).
It is then easy to
prove the commutativity
between the operators ${\hat\theta}^{\mu\nu}$,
whence they can be simultaneously diagonalized.
The DFR algebra defines a Lorentz-invariant
non-commutative space-time different from Snyder's
quantized space-time.
\\
\indent
Quite recently, 
Carlson, Carone and Zobin (CCZ) rederived\cite{5)} the DFR 
algebra\cite{4)}
by a certain contraction\footnote{CCZ's contraction
process is questionable
in view of the fact that
Snyder's coordinates
have a simple geometric
interpretation,
making it insufficient to
compare only an algebraic
similarity
between Snyder's and the DFR algebras.
In our opinion there is no connection between
the two algebras.}
of Snyder's algebra 
and formulated the non-commutative gauge theory (NCGT)
in a Lorentz-invariant way.
In their formulation of NCGT,
the old $\theta^{\mu\nu}$ becomes
an extra 6-dimensional coordinate
of the fields,
leading to an integration over it in the non-commutative action
which, however, contains no derivatives 
of the field quantities with respect to
the extra variables $\theta^{\mu\nu}$,
so that
the fields do not propagate into the $\theta$-space
of the hermitian operators, 
${\hat\theta}^{\mu\nu}$,
with no need of compactification
of the extra dimensions.
It turns out that
the constant $\theta$-algebra
violates the Lorentz invariance
because it singles out only one point
in the $\theta$-space,
the whole of which is needed to maintain
the Lorentz invariance.
The $\theta$-integration
is assumed to be controlled by a weight function
$W(\theta)$, 
which is normalized.
One (K. M.) of the authors applied\cite{6)} 
CCZ formulation to construct a realistic non-commutative 
QED and to derive an invariant damping factor.
\\
\indent
We continue in this paper our study on the Lorentz-invariant
non-commutative space-time
based on the DFR algebra.
The purpose of the present paper is four-fold.
The first is to present a simple way
of defining a non-commutative space-time
based on a representation
of the coordinate operators
in the momentum space.
Our derivation
of the DFR algebra
incidentally clarifies the reason that
the $c$-number $\theta$-algebra
is inconsistent theoretically.
The second is to
put CCZ formulation
into a firmer basis
by defining the operator fields
on the DFR algebra as operators diagonal
in the $\theta$-space.
Our method resorts to
a completeness relation
of the state vectors
obtained by diagonalizing
${\hat\theta}^{\mu\nu}$.
The third is to point out that
a general argument concerning
the normalizability
of the weight function
in the Lorentz metric
requires that
the $\theta$-space 
be
divided into
two disjoint (orthogonal) spaces
not connected by any Lorentz transformation.
CCZ covariant moment formula should
then be applied in each space, separately.
The fourth is to propose a 
non-commutative generalization
of Connes' $M_4\times Z_2$\cite{7)} in the standard model.
Two simple QFT models
are
reformulated on $M_4\times Z_2$
obtained in the commutative limit.
\\
\indent
The organization of this paper is as follows. In the next section
we argue that
the $c$-number $\theta$-algebra
is inconsistent from a theoretical point of view.
It is followed 
in \S3 by defining the DFR algebra
and the fields on it.
A comparison with Snyder's algebra
is also made.
The section 4 gives 
a general argument concerning
the normalizability
of the
weight function
in the Lorentz metric such
that
CCZ covariant moment formula 
should 
be applied in two disjoint spaces
not connected by any Lorentz
transformation, separately.
We propose a two-sheeted structure of the
Lorentz-invariant non-commutative space-time 
based on the DFR algebra
in \S5.
We formulate in \S6 two simple QFT models
on a discrete space-time
$M_4\times Z_2$
obtained in the commutative limit.
One of them relating the Higgs mechanism
to the discrete manifold \`a la Connes
was earlier discussed in a different
geometrical terminology.\cite{8)}
The last section is devoted to discussions.
We show in Appendix A that
the Lorentz generators
have different expressions for Snyder's and the DFR algebras.
\section{Inconsistency of $c$-number $\theta$-algebra} 
The canonical commutation relation
among the momentum and coordinate operators,\cite{9)}
\begin{eqnarray}
[{\hat p}^\mu,{\hat x}^\nu]&=&ig^{\mu\nu},
\label{eqn:2-1}
\end{eqnarray}
with {\it commuting} momentum variables,
\begin{eqnarray}
[{\hat p}^\mu,{\hat p}^\nu]&=&0,
\label{eqn:2-2}
\end{eqnarray}
does not necessarily imply that
the coordinate operators form an Abelian group of
the translations in the momentum space.
Instead we have, in general,
\begin{eqnarray}
{\hat x}^\mu=-i\frac{\partial}{\partial p_\mu}
+f^\mu(p),
\label{eqn:2-3}
\end{eqnarray}
in the $p$-space,
where
$f^\mu(p)$ is a 4-vector function of the momentum.
Then the commutator between the coordinate operators
become,
\begin{eqnarray}
[{\hat x}^\mu,{\hat x}^\nu]=-i\big(\frac{\partial f^\nu(p)}{\partial p_\mu}
-\frac{\partial f^\mu(p)}{\partial p_\nu}\big).
\label{eqn:2-4}
\end{eqnarray}
If the function $f^\mu(p)$ is given by a gradient,
\begin{eqnarray}
f^\mu(p)=\frac{\partial f(p)}{\partial p_\mu},
\label{eqn:2-5}
\end{eqnarray}
the right-hand side of (\ref{eqn:2-4})
vanishes so that the coordinate operators
commute with each other.
Conversely, if the coordinate operators
commute with each other,
the function $f^\mu(p)$ must be a gradient, and
the function $f(p)$ can be eliminated
through the redefinition of the
wave function, $\langle p|\psi\rangle
\to e^{-if(p)}\langle p|\psi\rangle$
up to a constant phase factor.
\\
\indent
We now suppose that $f^\mu(p)$
is no longer a gradient.
The simplest assumption is that
it contains no derivative with
respect to the momentum,
\begin{eqnarray}
f^\mu(p)=f^{\mu\nu}(p)p_\nu,
\label{eqn:2-6}
\end{eqnarray}
where $f^{\mu\nu}(p)$ is a 2-tensor.
There are two symmetric tensors available,
$p^\mu p^\nu$ and the metric tensor.
However, both
$p^\mu p^2$ and $p^\mu$ are pure gradients
and should be excluded from consideration
by the
assumption that $f^\mu(p)$
is not a gradient.
Consequently, we must assume
that $f^{\mu\nu}(p)$ is anti-symmetric.
Let $\theta^{\mu\nu}$ 
be an anti-symmetric
tensor
which can not depend on $p$. Then we have, putting
$f^{\mu\nu}=\theta^{\mu\nu}/2$,
\begin{eqnarray}
{\hat x}^\mu&=&-i\frac{\partial}{\partial p_\mu}
+\frac 12\theta^{\mu\nu}p_\nu,
\label{eqn:2-7}
\end{eqnarray}
which 
{\it formally} reproduces the algebra (\ref{eqn:1-1})
with the 2-tensor $\theta^{\mu\nu}$.
Unfortunately, however,
the Lorentz covariance
can not be maintained for the solution (\ref{eqn:2-7})
unless $\theta^{\mu\nu}=0$
despite of its appearance.
To see this we take
the commutator of ${\hat M}^{\mu\nu}$ 
with (\ref{eqn:1-1}),
which 
gives an identity,
\begin{eqnarray}
-\theta^{\mu\sigma}g^{\nu\rho}
+\theta^{\mu\rho}g^{\nu\sigma}+
\theta^{\nu\sigma}g^{\mu\rho}-
\theta^{\nu\rho}g^{\mu\sigma}=0.
\label{eqn:2-8}
\end{eqnarray}
Putting $\mu=\sigma\not=
\nu,\rho$ and using the anti-symmetry
of $\theta^{\mu\nu}$ we obtain,
\begin{eqnarray}
\theta^{\nu\rho}=0.
\label{eqn:2-9}
\end{eqnarray}
This result is obtained
as long as
the commutator between the coordinate
operators
is assumed to be a $c$-number
$\theta^{\mu\nu}$,
either regarded as a constant or a 2-tensor.
Consequently, the algebra
(\ref{eqn:1-1}) for the
nonvanishing $\theta^{\mu\nu}$
violates the Lorentz covariance.
Usually one ignores this difficulty,
looking for\cite{10)}
observable effects of
the Lorentz violation 
based on the constant $\theta$-algebra.\footnote{This 
Lorentz violation is to be contrasted
to a possible Lorentz violation considered in recent
literature\cite{11)} in which
small violations of the
Lorentz invariance
are described in terms of the Lorentz covariant fields.
On the contrary,
the constant $\theta$ algebra
does not allow us to define the Lorentz covariant fields
unless the Seiberg-Witten map\cite{12)}
is exploited. See the next section.} 
However, this point of view is not allowed theoretically.
Any non-commutative space-time should be 
characterized by
a {\it consistent} algebra which is closed under 
the commutation relations for the operators
$\{{\hat x}^\mu, {\hat M}^{\mu\nu}\}$.
There {\it is} an example (\ref{eqn:1-4})
that obeys this requirement.
The commutation relation
(\ref{eqn:1-1}) and the last two ones of
(\ref{eqn:1-4}) are closed.
But the commutator of ${\hat M}^{\mu\nu}$ with (\ref{eqn:1-1})
is inconsistent unless $\theta^{\mu\nu}=0$.
\section{DFR algebra and fields on it} 
Let us now try to restore the Lorentz covariance.
The next simplest case
would be to suppose that $\theta^{\mu\nu}$
in (\ref{eqn:2-7})
is regarded as an anti-symmetric 2-tensor
operator ${\hat\theta}^{\mu\nu}$,
\begin{eqnarray}
f^\mu(p)=\frac 12{\hat\theta}^{\mu\nu}p_\nu,
\label{eqn:3-1}
\end{eqnarray}
where
\begin{eqnarray}
[{\hat M}^{\mu\nu},{\hat\theta}^{\rho\sigma}]&=&i(g^{\nu\rho}{\hat\theta}^{\mu\sigma}
-g^{\nu\sigma}{\hat\theta}^{\mu\rho}-g^{\mu\rho}{\hat\theta}^{\nu\sigma}
+g^{\mu\sigma}{\hat\theta}^{\nu\rho}).
\label{eqn:3-2}
\end{eqnarray}
Because the operator ${\hat\theta}^{\mu\nu}$ is
independent of the momentum,
it commutes with the momentum operator,
\begin{eqnarray}
[{\hat\theta}^{\mu\nu}, {\hat p}^\rho]&=&0.
\label{eqn:3-3}
\end{eqnarray}
If we assume that the operator 
${\hat\theta}^{\mu\nu}$ commute with each other,
\begin{eqnarray}
[{\hat\theta}^{\mu\nu}, {\hat\theta}^{\rho\sigma}]&=&0,
\label{eqn:3-4}
\end{eqnarray}
we immediately find the commutator
\begin{eqnarray}
[{\hat x}^\mu,{\hat x}^\nu]&=&i{\hat\theta}^{\mu\nu}.
\label{eqn:3-5}
\end{eqnarray}
Conversely, if we assume (\ref{eqn:3-5})
together with the commutator,
\begin{eqnarray}
[{\hat\theta}^{\mu\nu}, {\hat x}^\rho]&=&0,
\label{eqn:3-6}
\end{eqnarray}
the commutator (\ref{eqn:3-4})
is obtained by the Jacobi
identity,
$[{\hat\theta}^{\mu\nu},[{\hat x}^\rho,{\hat x}^\sigma]]+$cyclic=0.
Similarly, the commutator (\ref{eqn:3-3})
is derived from the Jacobi identity,
$[{\hat p}^\mu,[{\hat x}^\nu,{\hat x}^\rho]]$+cyclic=0.
Any quantity that is independent
of the canonical variables
is not dynamical, so is not ${\hat\theta}^{\mu\nu}$.
\\
\indent
The commutation relations
(\ref{eqn:3-5}),
(\ref{eqn:3-6}) and (\ref{eqn:3-4})
define the DFR algebra.\cite{4)}
For this algebra
an inconsistency
discussed in the previous section
regarding the $c$-number
$\theta$-algebra (\ref{eqn:1-1})
is absolutely absent:
the commutator
of the Lorentz generators
${\hat M}^{\mu\nu}$
with (\ref{eqn:3-5})
only reproduces
(\ref{eqn:3-2}).
By construction the DFR algebra
is Lorentz-covariant
and closed under the commutation relations
for the operators
$\{{\hat x}^\mu, {\hat p}^\mu, {\hat\theta}^{\mu\nu},
{\hat M}^{\mu\nu}\}$.\footnote{The following commutator
should be added,
$[{\hat M}^{\mu\nu},{\hat p}^\rho]=i(g^{\nu\rho}{\hat p}^\mu
-g^{\mu\rho}{\hat p}^\nu)$.
}
\\
\indent
Our derivation of the DFR algebra
makes it easy to compare with
Snyder's algebra.
Let us now suppose that
the function $f^\mu(p)$ in (\ref{eqn:2-3})
is linear in the derivative, $\partial/\partial p$,
\begin{eqnarray}
f^\mu(p)=f^{\mu\nu}(p)\frac{\partial}{\partial p^\nu},
\label{eqn:3-7}
\end{eqnarray}
where $f^{\mu\nu}(p)$
is a 2-tensor function of $p$.
In this case the commutator 
$[{\hat p}^\mu,{\hat x}^\nu]$ gets no longer canonical,
\begin{eqnarray}
[{\hat p}^\mu,{\hat x}^\nu]&=&ig^{\mu\nu}
-f^{\nu\mu}.
\label{eqn:3-8}
\end{eqnarray}
This deviation
from the canonical commutation
relations
is only allowed if
the boundary condition that
the extra term $-f^{\nu\mu}$
in (\ref{eqn:3-8})
vanishes in the commutative limit
is satisfied.
Then the generalized Heisenberg
uncertainty relations
can be proved for (\ref{eqn:3-8})
(see, for instance, Ref.13)).
Because $f^{\mu\nu}(p)$
can not be constant,\footnote{There exists no
constant anti-symmetric tensor.
If
$f^{\mu\nu}(p)$ is a constant symmetric
tensor $g^{\mu\nu}$, the coordinate 
operators commute with each other.
We consider a contrary case in what follows.}
the simplest possibility is to take,
\begin{eqnarray}
f^{\mu\nu}(p)=ia^2p^\mu p^\nu,
\label{eqn:3-9}
\end{eqnarray}
where $a$ has a dimension of length.
The commutative limit
corresponds to $a\to 0$.
The coordinate operators are then given by,
\begin{eqnarray}
{\hat x}^\mu=-i\frac{\partial}{\partial p_\mu}
+ia^2p^\mu p^\nu\frac{\partial}{\partial p^\nu}.
\label{eqn:3-10}
\end{eqnarray}
This is nothing but a
geometrical expression for Snyder's coordinates
given by Kadyshevskii.\cite{3)}
In fact, we can recover Snyder's algebra
(\ref{eqn:1-4}) if we identify 
${\hat M}^{\mu\nu}$ with the Lorentz generators
in the $p$-space,
\begin{eqnarray}
{\hat M}^{\mu\nu}=i\big(p^\mu
\frac{\partial}{\partial p_\nu}
-p^\nu
\frac{\partial}{\partial p_\mu}\big).
\label{eqn:3-11}
\end{eqnarray}
The fundamental length
$a$ in (\ref{eqn:1-4})
equals
$a$ in (\ref{eqn:3-9}).
We also note that the commutator
$[{\hat p}^\mu,{\hat x}^\nu]$ given by
(\ref{eqn:3-8}) with (\ref{eqn:3-9}) 
is identical to that obtained by
Snyder.\cite{2)}
We thus find two 
notable differences 
between
Snyder's and the DFR algebras,
\begin{eqnarray}
[{\hat x}^\mu,[{\hat x}^\nu,{\hat x}^\rho]]&&
\left\{
       \begin{array}{l}
       =0\qquad ({\rm DFR}\;{\rm algebra}),\\
       \not =0\qquad ({\rm Snyder}{\mbox{'}}{\rm s}\;{\rm algebra}),\\
       \end{array}
       \right.\nonumber\\[2mm]
[{\hat p}^\mu,{\hat x}^\nu]&&
\left\{
       \begin{array}{l}
       =ig^{\mu\nu}\qquad ({\rm DFR}\;{\rm algebra}),\\
       \not =ig^{\mu\nu}\qquad ({\rm Snyder}{\mbox{'}}{\rm s}\;{\rm algebra}).\\
       \end{array}
       \right.
\label{eqn:3-12}
\end{eqnarray}
In what follows we exclusively employ the DFR 
algebra for its simpler nature,
although it contains a non-dynamical
quantity ${\hat\theta}^{\mu\nu}$ for the Lorentz invariance.
\\
\indent
Because of (\ref{eqn:3-4}) it is possible to diagonalize 
${\hat\theta}^{\mu\nu}$ simultaneously,
\begin{eqnarray}
{\hat\theta}^{\mu\nu}|\theta\rangle=\theta^{\mu\nu}|\theta\rangle,
\label{eqn:3-13}
\end{eqnarray}
where the eigenvalue $\theta^{\mu\nu}$ transforms as
an antisymmetric 2-tensor.
The states $|\theta\rangle$ are orthogonal for different
eigenvalues and normalized as follows,
\begin{eqnarray}
\langle\theta|\theta'\rangle=\delta^6(\theta-\theta')/W(\theta),
\label{eqn:3-14}
\end{eqnarray}
where $\delta^6(\theta-\theta')\equiv
\delta(\theta^{01}-\theta'{}^{01})\delta(\theta^{02}-\theta'{}^{02})
\delta(\theta^{03}-\theta'{}^{03})\delta(\theta^{12}-\theta'{}^{12})
\delta(\theta^{23}-\theta'{}^{23})\delta(\theta^{31}-\theta'{}^{31})$
and $W(\theta)$ is introduced for later purpose.
We also assume 
the completeness relation,
\begin{eqnarray}
\int\! d^6\theta W(\theta)
|\theta\rangle\langle\theta|=1.
\label{eqn:3-15}
\end{eqnarray}
\indent
Now the commutator (\ref{eqn:1-1}) is valid as a
`weak' relation
\begin{eqnarray}
\langle \theta|[{\hat x}^\mu,{\hat x}^\nu]|\theta\rangle
&=&i\theta^{\mu\nu}\langle \theta|\theta\rangle,
\label{eqn:3-16}
\end{eqnarray}
with due care of the unnormalizability of the state $|\theta\rangle$.
The quantity $\theta^{\mu\nu}$ here is a 2-tensor
specifying a point in the $\theta$-space,
the whole of which constitutes the Lorentz-invariant manifold.
On the contrary, the constant $\theta^{\mu\nu}$ in (\ref{eqn:1-1})
singles out one point in the $\theta$-space,
and,
hence, (\ref{eqn:1-1}) violates the Lorentz invariance
of the theory.
Even if the algebra (\ref{eqn:1-1}) is
considered to be Lorentz-covariant
as discussed below it,
it still violates the Lorentz invariance:
the Lorentz covariance
of the relation (\ref{eqn:3-16})
is derived from the Lorentz
transformation
property of the operator
${\hat\theta}^{\mu\nu}$ and, hence, the
state vector
$|\theta\rangle$
which is missing
in the $c$-number algebra. That is, 
the algebra (\ref{eqn:1-2}) should be regarded as a `weak' relation
of the commutator (\ref{eqn:3-5}) sandwiched between the 
Lorentz-transformed states,
$\langle \theta'|$ and $|\theta'\rangle$, with 
$\theta'=(\Lambda^{-1})^2\theta$.
\\
\indent
Now the Weyl representation of the DFR algebra is given by
the operators
\begin{eqnarray}
{\hat T}(p,\sigma)=
e^{ip_\mu{\hat x}^\mu+i\sigma_{\mu\nu}{\hat\theta}^{\mu\nu}}
\equiv
e^{ip{\hat x}+i\sigma{\hat\theta}},
\label{eqn:3-17}
\end{eqnarray}
with the multiplication law
\begin{eqnarray}
{\hat T}(p_1,\sigma_1){\hat T}(p_2,\sigma_2)
&=&
e^{-{\mbox{\tiny$\frac i2$}}(p_1\times p_2){\hat\theta}}
{\hat T}(p_1+p_2,\sigma_1+\sigma_2)\nonumber\\[2mm]
&=&{\hat T}(p_1+p_2,\sigma_1+\sigma_2-\frac 12p_1\times p_2).
\label{eqn:3-18}
\end{eqnarray}
Here we introduced the shorthand notation,
$(p_1\times p_2){\hat \theta}
=(p_1\times p_2)_{\mu\nu}{\hat \theta}^{\mu\nu}$
with $(p_1\times p_2)_{\mu\nu}\equiv
(1/2)(p_{1\mu}p_{2\nu}-p_{1\nu}p_{2\mu})$.
If we take the expectation value
of (\ref{eqn:3-18}) 
between the states $|\theta\rangle$
using the relations (\ref{eqn:3-4}),
(\ref{eqn:3-6}) and the completeness relation (\ref{eqn:3-15}), 
and put
\begin{eqnarray}
\langle \theta|{\hat T}(p,\sigma)
|\theta\rangle
&=&{\hat T}_\theta(p)e^{i\sigma\theta}
\langle \theta|\theta\rangle,\;\;\;
{\hat T}_\theta(p)=e^{ip{\hat x}},
\label{eqn:3-19}
\end{eqnarray}
we find the following multiplication law of the operators
${\hat T}_\theta(p)$,
\begin{eqnarray}
{\hat T}_\theta(p_1){\hat T}_\theta(p_2)
&=&
e^{-{\mbox{\tiny$\frac i2$}}(p_1\times p_2)\theta}
{\hat T}_\theta(p_1+p_2).
\label{eqn:3-20}
\end{eqnarray}
The subscript of the operators ${\hat T}_\theta(p)$
reminds of the `weak' relation (\ref{eqn:3-16}).
In this way we reproduce the multiplication law
of the operators ${\hat T}_\theta(p)$ for the $c$-number $\theta$-algebra
(\ref{eqn:1-1}).\cite{14)}
Consequently, if we define the operator field on the DFR algebra
in the Weyl representation, 
\begin{eqnarray}
{\hat \varphi}({\hat x},{\hat \theta})&=&\displaystyle{1\over (2\pi)^{4}}
\int\!d^4pd^6\sigma{\tilde\varphi}(p,\sigma){\hat T}(p,\sigma),
\label{eqn:3-21}
\end{eqnarray}
and take the the expectation value
between the states $|\theta\rangle$
using the relations (\ref{eqn:3-4}) and
(\ref{eqn:3-6}),
we obtain
\begin{eqnarray}
\langle \theta|{\hat \varphi}({\hat x},{\hat\theta})
|\theta\rangle
&=&{\hat \varphi}({\hat x},\theta)
\langle \theta|\theta\rangle,
\label{eqn:3-22}
\end{eqnarray}
where the operator field diagonal
in the $\theta$-space is put into the form,
\begin{eqnarray}
{\hat \varphi}({\hat x},\theta)&=&
\displaystyle{1\over (2\pi)^{4}}
\int\!d^4pd^6\sigma{\tilde\varphi}(p,\sigma)
e^{ip{\hat x}+i\sigma\theta}.
\label{eqn:3-23}
\end{eqnarray}
As usual, the Weyl symbol with respect to the operator
coordinates is defined by replacing them 
with the commuting coordinates,
\begin{eqnarray}
\varphi(x,\theta)&=&
\displaystyle{1\over (2\pi)^{4}}
\int\!d^4pd^6\sigma{\tilde\varphi}(p,\sigma)
e^{ipx+i\sigma\theta}.
\label{eqn:3-24}
\end{eqnarray}
Then, the product of operators
corresponds\cite{5)} to the Moyal $\ast$-product,
\begin{eqnarray}
{\hat \varphi}_1({\hat x},{\hat \theta})
{\hat \varphi}_2({\hat x},{\hat \theta})&=&{1\over (2\pi)^4}
\int\!d^4pd^6\sigma{\tilde \varphi}_{12}(p,\sigma)
e^{ip{\hat x}+i\sigma{\hat\theta}},\nonumber\\[2mm]
\varphi_{12}(x,\theta)&=&
{1\over (2\pi)^4}
\int\!d^4pd^6\sigma{\tilde \varphi}_{12}(p,\sigma)
e^{ipx+i\sigma\theta}\nonumber\\[2mm]
&=&e^{\frac i2\theta^{\mu\nu}\frac{\partial}{\partial x^\mu}
\frac{\partial}{\partial y^\nu}}
\varphi_1(x,\theta)\varphi_2(y,\theta)|_{x=y}
\equiv\varphi_1(x,\theta)*\varphi_2(x,\theta).
\label{eqn:3-25}
\end{eqnarray}
This correspondence needs no explicit proof\cite{6)}
in the present formalism.
As in deducing (\ref{eqn:3-20}) from (\ref{eqn:3-18})
through (\ref{eqn:3-19}) it is easy to prove that the product of operators
is also diagonal in the $\theta$-space:
\begin{eqnarray}
\langle \theta|{\hat \varphi}_1({\hat x},{\hat \theta})
{\hat \varphi}_2({\hat x},{\hat \theta})|\theta'\rangle
&=&{\hat \varphi}_1({\hat x},\theta)
{\hat \varphi}_2({\hat x},\theta)
\langle \theta|\theta'\rangle.
\label{eqn:3-26}
\end{eqnarray}
It then follows from (\ref{eqn:3-23}) that
the product
of the operators,
${\hat \varphi}_1({\hat x},\theta)
{\hat \varphi}_2({\hat x},\theta)$,
corresponds\cite{14)} to the Moyal product
with the deformation parameter, 
$\theta^{\mu\nu}$.
Only difference
lies in the additional
dependence of the operators
on the {\it same} $\theta^{\mu\nu}$.
This is observed by Carlson, Carone and Zobin\cite{5)}
without considering the $\theta$-space
spanned by the basis vectors $|\theta\rangle$. It
gives a logical basis for the $\theta$-expansion.
\cite{15),16)}
\\
\indent
As emphasized in Ref.~6)
the Lorentz-covariant operator fields
can be defined only if the underlying algebra
among the operator coordinates is Lorentz-covariant.
For instance, the condition for the scalar field
\begin{eqnarray}
\varphi'(x',\theta')&=&\varphi(x,\theta),
\label{eqn:3-27}
\end{eqnarray}
where $x'=\Lambda x$ and $\theta'=\Lambda^2\theta$,
is translated into that of the scalar operator field 
\begin{eqnarray}
{\hat\varphi}'({\hat x},{\hat\theta})&=&{\hat\varphi}({\hat x}',{\hat\theta}'),
\label{eqn:3-28}
\end{eqnarray}
where ${\hat x}'=\Lambda^{-1}{\hat x}$ 
and ${\hat\theta}'=(\Lambda^{-1})^2{\hat\theta}$,
and vice versa.
Conversely, if the operator
coordinates {\it do not}
obey the Lorentz-covariant algebra,
one is unable to define the Lorentz-covariant (operator) fields
unless the Seiberg-Witten map\cite{12)} is exploited,
which amounts to expressing the non-commutative fields
in terms of the Lorentz-covariant commutative fields
through the nonlinear field redefinition.
\\
\indent
Finally, we employ the trace property\cite{14)}
\begin{eqnarray}
{\rm tr}{\hat T}(p)=(2\pi)^4\delta^4(p),
\label{eqn:3-29}
\end{eqnarray}
and the completeness relation (\ref{eqn:3-15})
to obtain\footnote{The trace
in the $\theta$-space
is defined up to an irrelevant factor
as in (\ref{eqn:3-29}).
An explicit proof of the
trace formula (\ref{eqn:3-29})
with an irrelevant factor absorbed
can be seen, for instance,
in Appendix A of Ref.~15).} CCZ trace formula\cite{6)},
\begin{eqnarray}
{\rm tr}{\hat\varphi}({\hat x},{\hat\theta})=
\int\!d^4xd^6\theta W(\theta)\varphi(x,\theta).
\label{eqn:3-30}
\end{eqnarray}
A Lorentz-invariant
non-commutative action is then defined as in Ref.~6):
\begin{eqnarray}
{\hat S}=
\int\!d^4xd^6\theta W(\theta)
{\cal L}(\varphi(x,\theta),\partial_\mu
\varphi(x,\theta))_*,
\label{eqn:3-31}
\end{eqnarray}
where
the subscript
of the Lagrangian
means that
the $\ast$-product
should be taken for all
products
of the field variables.
It is to be recalled that,
although
the fields $\varphi(x,\theta)$ depend
on $x^\mu$ as well as $\theta^{\mu\nu}$,
the action thus constructed
contains the derivative with respect to
$x$ only,
so that
the fields do not propagate
into the extra $\theta$-space.
Hence we do not need any compactification
of the extra dimensions.
\section{Modified completeness relation
in the $\theta$-space}
We now investigate the completeness relation
(\ref{eqn:3-15}) in more detail.
We slightly change the previous notation
by noting the fact that
the commutator $[{\hat x}^\mu,{\hat x}^\nu]$
should vanish in the commutative limit,
$a\to 0$, 
where $a$ is a fundamental
length in the theory as in the case of Snyder's quantized
space-time.\footnote{At the present time
it is still optional to choose the non-commutative scale
so that we are free to determine it in this paper
provided no conflict with experiment
arises.}
By dimensional argument we put
\begin{eqnarray}
{\hat\theta}^{\mu\nu}=a^2{\hat{\bar\theta}}^{\mu\nu},
\label{eqn:4-1}
\end{eqnarray}
where ${\hat{\bar\theta}}^{\mu\nu}$ is a dimensionless
operator.
We write the state vector as
\begin{eqnarray}
{\hat{\bar\theta}}^{\mu\nu}|{\bar\theta}\rangle
={\bar\theta}^{\mu\nu}|{\bar\theta}\rangle,
\label{eqn:4-2}
\end{eqnarray}
so that
\begin{eqnarray}
{\hat\theta}^{\mu\nu}|{\bar\theta}\rangle
=\theta^{\mu\nu}|{\bar\theta}\rangle,
\label{eqn:4-3}
\end{eqnarray}
with\cite{6)}
\begin{eqnarray}
\theta^{\mu\nu}=a^2{\bar\theta}^{\mu\nu}.
\label{eqn:4-4}
\end{eqnarray}
Since $a$ is Lorentz scalar,
${\bar\theta}^{\mu\nu}$ is a
dimensionless anti-symmetric 2-tensor.
\\
\indent
The normalization condition (\ref{eqn:3-14}) reads
\begin{eqnarray}
\langle{\bar\theta}|{\bar\theta}'\rangle
=\delta^6({\bar\theta}-{\bar\theta}')/w({\bar\theta}),
\label{eqn:4-5}
\end{eqnarray}
where we put
\begin{eqnarray}
W(\theta)=a^{-12}w({\bar\theta}).
\label{eqn:4-6}
\end{eqnarray}
The completeness condition (\ref{eqn:3-15}) then becomes,
\begin{eqnarray}
\int\!d^6{\bar\theta}w({\bar\theta})|{\bar\theta}\rangle
\langle{\bar\theta}|
=1.
\label{eqn:4-7}
\end{eqnarray}
\indent
We also note that
the non-commutative action (\ref{eqn:3-31})
can be expanded in terms of the parameter
$a$ which is considered a fundamental length
in the theory:
\begin{equation}
{\hat S}=
S^{(0)}+S^{(2)}+S^{(4)}+\cdots,
\label{eqn:4-8}
\end{equation}
where $S^{(2n)}$
is of order $a^{2n}$, $n=0,1,2,\dots$.
We do not consider any question
of the convergence property
of this expansion.
Through
(\ref{eqn:4-4}) and the $\theta$-integration
in (\ref{eqn:3-31})
this expansion
is obtained from the $\theta$-expansion\cite{15),16)}
of the Lagrangian\footnote{In the old version
based on (\ref{eqn:1-1})
the $\theta$-expansion of the Lagrangian
is the same as that
of the non-commutative action,
but in the CCZ Lorentz-invariant
formulation
the $\theta$-expansion defined
for the Lagrangian
turns into the $a$-expansion
of the non-commutative action via the $\theta$-integration.}
\begin{eqnarray}
{\cal L}(\varphi(x,\theta),\partial_\mu
\varphi(x,\theta))_*
={\cal L}^{(0)}+{\cal L}^{(1)}+{\cal L}^{(2)}+\cdots,
\label{eqn:4-9}
\end{eqnarray}
where ${\cal L}^{(n)}$
is of order $n$ in $\theta^{\mu\nu}$,
$n=0,1,2,\dots$.
The reason that
the odd terms ${\cal L}^{(2n+1)}$
disappear in the expansion
(\ref{eqn:4-8}) stems from the fact that
$W(\theta)$
is an even function:
\begin{eqnarray}
W(-\theta)=W(\theta).
\label{eqn:4-10}
\end{eqnarray}
This property\cite{5)} is
obvious
from the Lorentz invariance
of $W(\theta)$.
We recover QFT action
in the commutative limit, $a\to 0$,
\begin{eqnarray}
S\equiv S^{(0)}=\int\!d^4x
{\cal L}^{(0)}(\varphi(x),\partial_\mu
\varphi(x))\equiv
\int\!d^4x{\cal L}(\varphi(x),\partial_\mu
\varphi(x)),
\label{eqn:4-11}
\end{eqnarray}
provided 
\begin{eqnarray}
\int\!d^6\theta W(\theta)=1.
\label{eqn:4-12}
\end{eqnarray}
This is
CCZ normalization condition
which is equivalent to,
\begin{eqnarray}
\int\!d^6{\bar\theta} w({\bar\theta})=1.
\label{eqn:4-13}
\end{eqnarray}
\indent
Dynamical content
of the weight function
is revealed through
the substitution of
the expansion (\ref{eqn:4-9})
into the definition
(\ref{eqn:3-31}).
Neglecting the 
odd-order terms in (\ref{eqn:4-9})
as explained above,
we encounter the integrals,
\begin{eqnarray}
\int\!d^6{\bar\theta} w({\bar\theta})
\prod_{i=1}^{2n}{\bar\theta}^{\mu_i\nu_i},\quad n=0,1,2,\cdots.
\label{eqn:4-14}
\end{eqnarray}
These integrals can be
expressed in terms of the invariant moments,
\begin{eqnarray}
\langle{\bar\theta}^{2n}
\rangle=\int\!d^6{\bar\theta}w({\bar\theta})
({\bar\theta}^{\mu\nu}{\bar\theta}_{\mu\nu})^n, n=0,1,2,\cdots.
\label{eqn:4-15}
\end{eqnarray}
The normalization
condition (\ref{eqn:4-12})
implies $\langle 1\rangle=1$, corresponding to the case,
$n=0$.
For $n=1$, we have\cite{5)}
\begin{eqnarray}
\int\!d^6{\bar\theta}w({\bar\theta}){\bar\theta}^{\mu_1\nu_1}
{\bar\theta}^{\mu_2\nu_2}&=&\frac {\langle {\bar\theta}^2\rangle}{12}
(g^{\mu_1\mu_2}g^{\nu_1\nu_2}
-g^{\mu_1\nu_2}g^{\nu_1\mu_2}),
\label{eqn:4-16}
\end{eqnarray}
for $n=2$, we get\cite{6)}
\begin{eqnarray}
\int\!d^6{\bar\theta}w({\bar\theta}){\bar\theta}^{\mu_1\nu_1}
{\bar\theta}^{\mu_2\nu_2}{\bar\theta}^{\mu_3\nu_3}
{\bar\theta}^{\mu_4\nu_4}
\nonumber\\[2mm]
&&\!\!\!\!\!\!\!\!\!\!\!\!\!\!\!\!\!\!\!\!
\!\!\!\!\!\!\!\!\!\!\!\!\!\!\!\!\!\!\!\!=
\frac{\langle{\bar\theta}^4\rangle}{192}
[(g^{\mu_1\mu_2}g^{\nu_1\nu_2}-g^{\mu_1\nu_2}g^{\nu_1\mu_2})
(g^{\mu_3\mu_4}g^{\nu_3\nu_4}-g^{\mu_3\nu_4}g^{\nu_3\mu_4})
\nonumber\\[2mm]
&&\!\!\!\!\!\!\!\!\!\!\!\!\!\!\!\!\!\!\!\!
\!\!\!\!\!\!\!\!\!\!\!\!\!\!\!\!\!\!\!\!
+(g^{\mu_1\mu_3}g^{\nu_1\nu_3}-g^{\mu_1\nu_3}g^{\nu_1\mu_3})
(g^{\mu_2\mu_4}g^{\nu_2\nu_4}-g^{\mu_2\nu_4}g^{\nu_2\mu_4})
\nonumber\\[2mm]
&&\!\!\!\!\!\!\!\!\!\!\!\!\!\!\!\!\!\!\!\!
\!\!\!\!\!\!\!\!\!\!\!\!\!\!\!\!\!\!\!\!
+(g^{\mu_1\mu_4}g^{\nu_1\nu_4}-g^{\mu_1\nu_4}g^{\nu_1\mu_4})
(g^{\mu_2\mu_3}g^{\nu_2\nu_3}-g^{\mu_2\nu_3}g^{\nu_2\mu_3})],
\label{eqn:4-17}
\end{eqnarray}
and so on.
\\
\indent
We now evaluate
the integral (\ref{eqn:4-15})
by 
assuming that
the weight function
$w({\bar\theta})$
is a function
of only the invariant
\begin{eqnarray}
{\bar\alpha}&\equiv&
\frac 12{\bar\theta}^{\mu\nu}{\bar\theta}_{\mu\nu}.
\label{eqn:4-18}
\end{eqnarray}
Using the parametrization 
\begin{eqnarray}
{\bar\theta}^{01}&=&\rho\sin{\theta}\cos{\phi},\;\;\;
{\bar\theta}^{02}=\rho\sin{\theta}\sin{\phi},\;\;\;
{\bar\theta}^{03}=\rho\cos{\theta},\nonumber\\[2mm]
{\bar\theta}^{12}&=&\sigma\sin{\vartheta}\cos{\varphi},\;\;\;
{\bar\theta}^{23}=\sigma\sin{\vartheta}\sin{\varphi},\;\;\;
{\bar\theta}^{31}=\sigma\cos{\vartheta},
\label{eqn:4-19}
\end{eqnarray}
we find
\begin{eqnarray}
\langle{\bar\theta}^{2n}
\rangle
=16\pi^2\int_0^\infty\!\rho^2d\rho
\int_0^\infty\!\sigma^2d\sigma 
w(-\rho^2+\sigma^2)(-\rho^2+\sigma^2)^n.
\label{eqn:4-20}
\end{eqnarray}
For this integral to converge for
all $n$
the function $w(-\rho^2+\sigma^2)$
must act as an exponential damping factor.
Because of the negative sign
in front of $\rho^2$,
which is unavoidable in the Lorentz metric,
any exponential damping factor
becomes powerless
on the line
$\rho=\sigma$,
which corresponds to the light-like 
region ${\bar\alpha}=0$. 
To see this more explicitly
we put,
\begin{eqnarray}
\rho&=&r\cos{\psi},\nonumber\\[2mm]
\sigma&=&r\sin{\psi},
\label{eqn:4-21}
\end{eqnarray}
so that the region ${\bar\alpha}>0$ corresponds to
\begin{eqnarray}
0<r<\infty,\;\;\;\psi=\frac \pi 4+\epsilon\to \psi=\frac \pi2,
\label{eqn:4-22}
\end{eqnarray}
while the region ${\bar\alpha}<0$ corresponds to
\begin{eqnarray}
0<r<\infty,\;\;\;\psi=0\to \psi=\frac \pi4-\epsilon.
\label{eqn:4-23}
\end{eqnarray}
The equation $\psi=\frac \pi4$ defines
the line $\rho=\sigma$ to be excluded
from the integration region.
Consequently, the parameter $\epsilon$ is to be chosen positive.
For ${\bar\alpha}>0$ we get
\begin{eqnarray}
\langle{\bar\theta}^{2n}
\rangle_+&\equiv&
4\pi^2\int_0^\infty\!r^{(5+2n)}dr\int_{\pi/4+\epsilon}^{\pi/2}
d\psi(-\cos{2\psi})^n(\sin^2{2\psi})w(-r^2\cos{2\psi})
\nonumber\\[2mm]
&=&\pi^2B_{1-\epsilon}(\frac 32,-1)
\int_0^\infty dRR^{5+2n}w(R^2),
\label{eqn:4-24}
\end{eqnarray}
where
$R^2=-r^2\cos{2\psi}$ is a Lorentz-invariant
variable and,
\begin{eqnarray}
B_x(p,q)=\int_0^xdtt^{p-1}(1-t)^{q-1},
\label{eqn:4-25}
\end{eqnarray}
is the incomplete beta function.
Similarly, we obtain
\begin{eqnarray}
\langle{\bar\theta}^{2n}
\rangle_-&\equiv&
4\pi^2\int_0^\infty\!r^{(5+2n)}dr\int_0^{\pi/4-\epsilon}
d\psi(-\cos{2\psi})^n(\sin^2{2\psi})w(-r^2\cos{2\psi})
\nonumber\\[2mm]
&=&
\pi^2B_{1-\epsilon}(\frac 32,-1)
\int_0^\infty dRR^{5+2n}(-1)^nw(-R^2).
\label{eqn:4-26}
\end{eqnarray}
Here $R^2=r^2\cos{2\psi}$.
In both expressions
(\ref{eqn:4-24}) and (\ref{eqn:4-26})
$R^2$ is positive.
The divergent factor
$B_{1-\epsilon}(\frac 32,-1)$,
which, at first sight, seems to violate
the Lorentz invariance,
can be eliminated through the normalization condition,
\begin{eqnarray}
\langle 1\rangle&=&
\langle 1\rangle_++\langle 1\rangle_-\nonumber\\[2mm]
&=&
\pi^2B_{1-\epsilon}(\frac 32,-1)
\int_0^\infty dRR^5\big(w(R^2)+w(-R^2)\big)=1.
\label{eqn:4-27}
\end{eqnarray}
Then one can express
all the moments
as,
\begin{eqnarray}
\langle{\bar\theta}^{2n}
\rangle&=&
\langle{\bar\theta}^{2n}
\rangle_++
\langle{\bar\theta}^{2n}
\rangle_-\nonumber\\[2mm]
&=&\frac
{\int_0^\infty dRR^{5+2n}
\big(w(R^2)+(-1)^nw(-R^2)\big)}
{\int_0^\infty dRR^5\big(w(R^2)+w(-R^2)\big)}.
\label{eqn:4-28}
\end{eqnarray}
\indent
This integral converges only if both $w(R^2)$ and $w(-R^2)$
diminishes exponentially at $R^2\to\infty$. In general, however,
this is not the case. For instance, if we put
\begin{eqnarray}
w({\bar\theta})=Ne^{-b{\bar\alpha}^2+c{\bar\alpha}}
=Ne^{-bR^4+cR^2}\equiv w(R^2),
\label{eqn:4-29}
\end{eqnarray}
there are two possible choices, $b>0$ or $b=0$.\footnote{For $b>0$,
it may happen that $c=0$. In this case $w(R^2)$ is an even function
of $R^2$ so that the invariant moments for $n=1,3,\cdots$
identically vanish. This gives rise to a contradiction
because, in general, the left-hand side of (\ref{eqn:4-16})
does not vanish componentwise. This is another reason that
we are led to the following Lorentz-invariant decomposition
of the $\theta$-space.}
For $b>0$, both $w(R^2)$ and $w(-R^2)$ go to 0 at $R^2\to\infty$,
while, for $b=0$, $w(R^2)$ and $w(-R^2)$ vanish exponentially 
at $R^2\to\infty$ if $c<0$ and $c>0$, respectively.
Namely, we have, in general, to choose the weight function separately
according to the sign of the invariant ${\bar\alpha}$.
\\
\indent
This is in accord with the following situation.
Since the two regions
with different sign of the invariant
${\bar\alpha}$
can not be connected
by any (including improper)
Lorentz transformation,
one can define
the field variable
$\varphi(x,\theta)$
independently
in the two regions,
denoted $\varphi_\pm(x,\theta)$,
where $\epsilon=\pm$ corresponds to
positive
and negative
${\bar\alpha}$, respectively.
There are no Lorentz-invariant
interactions
between the
fields $\varphi_\epsilon(x,\theta)$
for different $\epsilon=\pm$.
This suggests a division
of the $\theta$-space
into two disjoint
spaces according to the sign of
${\bar\alpha}$.
This Lorentz-invariant 
decomposition
precisely corresponds to the necessity of the separate choice of the weight
function according to the sign of ${\bar\alpha}$.
Consequently, we have to define the invariant moments
separately in each disjoint space,
\begin{eqnarray}
\langle{\bar\theta}^{2n}
\rangle_\pm&\equiv&
\int_\pm\!d^6{\bar\theta}
w_\pm({\bar\theta}),
\label{eqn:4-30}
\end{eqnarray}
where the integral
$\int_{\pm}$
denotes that the integration
region is restricted to
positive ${\bar\alpha}$
and
negative ${\bar\alpha}$ as in (\ref{eqn:4-24}) and 
(\ref{eqn:4-26}), respectively,
while
the weight function
$w_\pm(\pm R^2)$
vanishes exponentially
at large $R^2$.
It turns out
that
\begin{eqnarray}
\langle{\bar\theta}^{2n}
\rangle_+&=&
\frac
{\int_0^\infty dRR^{5+2n}
w_+(R^2)}
{\int_0^\infty dRR^5w_+(R^2)},\nonumber\\[2mm]
\langle{\bar\theta}^{2n}
\rangle_-&=&
(-1)^n\frac
{\int_0^\infty dRR^{5+2n}
w_-(-R^2)}
{\int_0^\infty dRR^5w_-(-R^2)},
\label{eqn:4-31}
\end{eqnarray}
where we assumed $\langle 1\rangle_+=
\langle 1\rangle_-=1$.
Also, 
the covariant moment formula
(\ref{eqn:4-16})
is modified as
\begin{eqnarray}
\int_\pm\!d^6{\bar\theta}w_\pm({\bar\theta}){\bar\theta}^{\mu_1\nu_1}
{\bar\theta}^{\mu_2\nu_2}&=&
\frac {\langle {\bar\theta}^2\rangle_\pm}{12}
(g^{\mu_1\mu_2}g^{\nu_1\nu_2}
-g^{\mu_1\nu_2}g^{\nu_1\mu_2}),
\label{eqn:4-32}
\end{eqnarray}
and similarly for the higher-order covariant moment formulae.
\\
\indent
According to
the division of the $\theta$-space
into
two disjoint spaces not
connected by any Lorentz
transformation,
we write the state vectors as
$|{\bar\theta}\rangle_\pm$,
where the sign $\pm$ indicates the signature of
the invariant ${\bar\alpha}$.
The completeness condition (\ref{eqn:3-15})
then reads
\begin{eqnarray}
\sum_{\epsilon=\pm}
\int_\epsilon\!d^6{\bar\theta}w({\bar\theta})
|{\bar\theta}\rangle_{\!\epsilon}\;
{}_{\epsilon\!}\langle{\bar\theta}|
=1.
\label{eqn:4-33}
\end{eqnarray}
Hence, the state vectors of different signatures
are orthogonal,\footnote{The delta function
$\delta^6({\bar\theta}-{\bar\theta}')$
in the equation below already implies the
Kronecker delta $\delta_{\epsilon\epsilon'}$
in the same equation.
For later convenience
we write both deltas.}
\begin{eqnarray}
{}_{\epsilon}\langle{\bar\theta}|
{\bar\theta}'\rangle_{\epsilon'}
=\delta^6({\bar\theta}-{\bar\theta}')
\delta_{\epsilon\epsilon'}/w({\bar\theta}).
\label{eqn:4-34}
\end{eqnarray}
Because the states $|{\bar\theta}\rangle_\pm$
are not connected by
any Lorentz transformation
and are orthogonal to each other,
the expectation values
of the operator
fields are diagonal also in the signatures,
\begin{eqnarray}
{}_\epsilon\langle {\bar\theta}|{\hat \varphi}({\hat x},{\hat\theta})
|{\bar\theta}'\rangle_{\epsilon'}
&=&{\hat \varphi}_\epsilon({\hat x},\theta)
\delta_{\epsilon{\epsilon'}}\;
{}_{\epsilon\,}\langle {\bar\theta}|{\bar\theta}'\rangle_\epsilon,
\label{eqn:4-35}
\end{eqnarray}
where $\epsilon, \epsilon'=\pm$.
This implies no interactions
between the fields
$\varphi_\epsilon(x,\theta)$
for different $\epsilon$.
Hence it is a matter of convention
to choose the plus sign or the minus sign.
\section{Non-commutative generalization
of Connes' space-time $M_4\times Z_2$}
Let us suppose that we live in the
space of positive ${\bar\alpha}$
and drop the plus sign attached to the state vector,
$|{\bar\theta}\rangle\equiv |{\bar\theta}\rangle_+$,
and the operator field
${\hat \varphi}({\hat x},\theta)
\equiv{\hat \varphi}_+({\hat x},\theta)$.
\\
\indent
In connection with the Lorentz-invariant non-commutative
space-time based on the DFR algebra we recall that
Connes\cite{7)}
assumed a discrete space-time $M_4\times Z_2$
to be the underlying space-time behind
the standard model.
Without entering into details,
Connes' idea is the following.
The standard model is a spontaneously
broken gauge theory,
containing 
matter, gauge and Higgs fields.
Since the existence of the gauge fields have a
geometrical interpretation,
the Higgs fields
might also have a geometric origin.
Although the gauge fields are defined on
a continuous manifold,
one can not introduce the Higgs fields
{\it as a kind of gauge fields}
on the same continuous manifold,
because the gauge fields
are related to
the propagating nature of the matter fields,
while the Higgs fields
are introduced into the theory
in relation to the mass term of fermions,
which contains no derivative.
Consequently,
a radical change of the
space-time geometry
is needed to understand
that all force-mediating fields may be
interpreted as originating
from the underlying geometry.
A concrete space-time structure
proposed by Connes\cite{7)}
is $M_4\times Z_2$
which is a discrete manifold.\footnote{Without
modifying the
underlying space-time,
Sogami\cite{17)} proposed to unify the gauge and the Higgs
fields in terms of the generalized covariant
derivative
acting on the total fermion field
including the chiral leptons and quarks.}
In this space the Higgs fields
can be interpreted as a gauge field
in the discrete direction,
while
the ordinary gauge fields
are defined on the continuous manifolds,
two copies of the Minkowski space-time.
The number 2 is related to
the dichotomic nature of the chirality of the fermions.
\\
\indent
It is then natural to ask if
Connes' idea
works also in the non-commutative
space-time.
We affirmatively answer this question
in this section.
To this purpose we introduce
the additional quantum number $y=\pm$
to characterize
the state $|{\bar\theta}\rangle$,
which is then denoted,
\begin{eqnarray}
|{\bar\theta},y\rangle,\quad
y=\pm.
\label{eqn:5-1}
\end{eqnarray}
The two states, $|{\bar\theta},\pm\rangle$,
for the {\it same} ${\bar\theta}$
are further assumed to be
non-orthogonal.
This assumption 
means that
there exist interactions between
the diagonal fields defined by the expectation value,
\begin{eqnarray}
\langle{\bar\theta},y|\varphi({\hat x},{\hat\theta})
|{\bar\theta}',y\rangle
=\varphi({\hat x},\theta,y,y)
\langle{\bar\theta},y|{\bar\theta}',y\rangle
\equiv \varphi({\hat x},\theta,y)
\langle{\bar\theta},y|{\bar\theta}',y\rangle,
\;\;y=\pm.
\label{eqn:5-2}
\end{eqnarray}
Such interactions must be mediated by the 
off-diagonal fields,
\begin{eqnarray}
\langle{\bar\theta},y|\varphi({\hat x},{\hat\theta})
|{\bar\theta}',-y\rangle
=\varphi({\hat x},\theta,y,-y)
\langle{\bar\theta},y|{\bar\theta}',-y\rangle,
\;y=\pm.
\label{eqn:5-3}
\end{eqnarray}
\indent
According to our assumption, any relativistic field
defined on the non-commutative space-time
based on the DFR algebra
has the following $c$-number
structure,
\begin{eqnarray}
{\rm monogenic}\;{\rm fields}&\;\;\;&
\varphi_\pm(x,\theta,y),\;\;\;\;\;\;\;\;\;y=\pm,\nonumber\\[2mm]
{\rm dichotomic}\;{\rm fields}
&\;\;\;&\varphi_\pm(x,\theta,y,-y),\;\;\;
y=\pm.
\label{eqn:5-4}
\end{eqnarray}
In the commutative limit, $a\to 0$,
we are left with the local fields,
$\varphi(x,y)=\varphi(x,\theta=0,y)$,
which are
supposed to describe the 
chiral matter fields and
the chiral gauge fields,
and the local fields 
$\varphi(x,y,-y)\equiv\varphi(x,\theta=0,y,-y)$,
which
are assumed to be
the Higgs fields
responsible for the masses of the
monogenic fields.
Consequently, the Higgs fields
on our non-commutative space-time
have a geometric origin
as a kind of the non-commutative gauge fields.
A detailed model construction
will be postponed to later publications.
In the next section
we present a plausible
argument to support this
interpretation in the commutative limit based
on our earlier work.\cite{8)}
\\
\indent
The same discussion is also valid for
the case, ${\bar\alpha}<0$.
\section{QFT on $M_4\times Z_2$}
As alluded to above,
to make our presentation as simple as possible
we restrict ourselves to the commutative limit
$a\to 0$ in this section,\footnote{Hence, there is no
$\ast$-product in what follows.}
still working in the space with
positive ${\bar\alpha}$.
\\
\indent
The free Dirac action on $M_4\times Z_2$
is given by,
\begin{eqnarray}
S_{D_0}=\sum_{y=\pm}
\int\!d^4x{\bar\psi}(x,y)(i\gamma^\mu\partial_\mu
+i\gamma^5\partial_y)\psi(x,y),
\label{eqn:6-1}
\end{eqnarray}
where\footnote{The matrix
$\gamma^5=i\gamma^0\gamma^1\gamma^2\gamma^3$
is inserted for later convenience.}we define the derivative
in the discrete direction as,
\begin{eqnarray}
\partial_y\psi(x,y)=M(y,-y)\psi(x,-y).
\label{eqn:6-2}
\end{eqnarray}
It should be noted that
the derivative
in the discrete direction
is required in relation with the mass
term of fermion,
which has no continuous derivative.
If we assume the hermiticity condition,
\begin{eqnarray}
M^{\dag}(y,-y)=M(-y,y),
\label{eqn:6-3}
\end{eqnarray}
the action (\ref{eqn:6-1})
is hermitian.
We
assign
the left-handed and the right-handed spinors
on the different sheets,
\begin{eqnarray}
\psi_L(x)=\psi(x,+),\qquad\psi_R(x)=\psi(x,-).
\label{eqn:6-4}
\end{eqnarray}
Upon transforming
$\psi_L(x)\to e^{i\pi/4}\psi_L(x)$
and
$\psi_R(x)\to e^{-i\pi/4}\psi_R(x)$,
we recover the well-known free Dirac action,
\begin{eqnarray}
S_{D_0}=
\int\!d^4x\big({\bar\psi}(x)i\gamma^\mu\partial_\mu\psi(x)
-{\bar\psi}(x)M\psi(x)\big),\;\;\;M\equiv M(+,-).
\label{eqn:6-5}
\end{eqnarray}
\indent
The gauge transformation
may or may not depend
on the variable $y$.
Let us first consider the following
gauge transformation independent of $y$,
\begin{eqnarray}
\psi(x,y)\to {}^g\psi(x,y)=g(x)\psi(x,y),
\label{eqn:6-6}
\end{eqnarray}
where $g(x)$ is unitary.
The gauge-invariant Dirac action
$S_D$
is obtained by the replacement,
\begin{eqnarray}
\partial_\mu&\to& D_\mu=\partial_\mu
+A_\mu(x),
\label{eqn:6-7}
\end{eqnarray}
in (\ref{eqn:6-1}),
with the gauge transformation property
\begin{eqnarray}
A_\mu(x)\to
{}^gA_\mu(x)=g(x)A_\mu(x)g^{\dag}(x)+g(x)\partial_\mu g^{\dag}(x).
\label{eqn:6-8}
\end{eqnarray}
The derivative $\partial_y$
is unchanged
because $\partial_yg(x)=0$,
the fermion mass being gauge-invariant.
Writing the gauge field as
\begin{eqnarray}
A(x)=i\gamma^\mu A_\mu(x),
\label{eqn:6-9}
\end{eqnarray}
we define the exterior derivative by
\begin{eqnarray}
dA(x)=\frac 12i\gamma^\mu\wedge i\gamma^\nu 
(\partial_\mu A_\nu(x)
-\partial_\nu A_\mu(x)),
\label{eqn:6-10}
\end{eqnarray}
and the wedge product by
\begin{eqnarray}
A(x)\wedge A(x)=
\frac 12i\gamma^\mu\wedge i\gamma^\nu
[A_\mu(x),A_\nu(x)],
\label{eqn:6-11}
\end{eqnarray}
to obtain the field strength
through
\begin{eqnarray}
F(x)&=&dA(x)+A(x)\wedge A(x)
=\frac 12i\gamma^\mu\wedge i\gamma^\nu
F_{\mu\nu}(x),\nonumber\\[2mm]
F_{\mu\nu}(x)&=&\partial_\mu A_\nu(x)
-\partial_\nu A_\mu(x)+[A_\mu(x),A_\nu(x)].
\label{eqn:6-12}
\end{eqnarray}
Here the wedge product
of $\gamma$ matrices is defined by
\begin{eqnarray}
\gamma^\mu\wedge\gamma^\nu
=\displaystyle{{1\over 2}}
(\gamma^\mu\gamma^\nu-\gamma^\nu\gamma^\mu)
\equiv -i\sigma^{\mu\nu}.
\label{eqn:6-13}
\end{eqnarray}
The bosonic action identical to the YM action
is then given by
\begin{eqnarray}
S_B=-\frac 18\int\!d^4x{\rm Tr}\frac 1{g^2}
F^2\,(x),
\label{eqn:6-14}
\end{eqnarray}
where $g^2$ is a matrix of the
coupling constants squared,
Tr includes both the traces
over the Dirac matrices and the internal symmetry
matrices. 
For the gauge group $U(1)$
the sum $S_D+S_B$
defines spinor QED.
\\
\indent
Let us next consider the gauge transformation
depending on the variable $y$,
\begin{eqnarray}
\psi(x,y)\to {}^g\psi(x,y)=g(x,y)\psi(x,y),
\label{eqn:6-15}
\end{eqnarray}
where $g(x,y)$ is unitary.\footnote{The differential geometry
on $M_4\times Z_2$
was developed in Ref.~8)
and we can use it for the case under consideration.
The following presentation
is only a translation of the result in Ref.~8).}
The gauge-invariant Dirac action
$S_{{\mbf D}}$
is obtained by the replacement,
\begin{eqnarray}
\partial_\mu&\to& D_\mu=\partial_\mu
+A_\mu(x,y),\nonumber\\[2mm]
\partial_y&\to& D_y=\partial_y
+\Phi(x,y,-y),
\label{eqn:6-16}
\end{eqnarray}
in (\ref{eqn:6-1}),
where the gauge field
$\Phi(x,y,-y)$
is assumed to flip the sign of $y$
of the spinor on the right,
\begin{eqnarray}
D_y\psi(x,y)=
\partial_y\psi(x,y)+
\Phi(x,y,-y)\psi(x,-y),
\label{eqn:6-17}
\end{eqnarray}
in accordance with the definition
(\ref{eqn:6-2}) of the derivative $\partial_y$.
The gauge transformation properties
of the gauge fields
are obtained as follows,
\begin{eqnarray}
A_\mu(x,y)&\to&
{}^gA_\mu(x,y)=g(x,y)A_\mu(x,y)g^{\dag}(x,y)
+g(x,y)\partial_\mu g^{\dag}(x,y),\nonumber\\[2mm]
\Phi(x,y,-y)&\to&
{}^g\Phi(x,y,-y)=g(x,y)\Phi(x,y,-y)g^{\dag}(x,-y)
-(\partial_yg(x,y))g^{\dag}(x,-y).
\label{eqn:6-18}
\end{eqnarray}
The derivative $\partial_yg(x,y)$
can be determined by the Leibniz rule,
\begin{eqnarray}
\partial_y(g(x,y)\psi(x,y))&=&
M(y,-y)g(x,-y)\psi(x,-y)\nonumber\\[2mm]
&=&
g(x,y)\partial_y\psi(x,y)
+(\partial_yg(x,y))\psi(x,-y),
\label{eqn:6-19}
\end{eqnarray}
where the first equality comes from the definition
(\ref{eqn:6-2}) and
the second equality
is the Leibniz rule.
Using (\ref{eqn:6-2})
we obtain
\begin{eqnarray}
\partial_yg(x,y)
=M(y,-y)g(x,-y)-g(x,y)M(y,-y).
\label{eqn:6-20}
\end{eqnarray}
Inserting this equation into
the second equation
of (\ref{eqn:6-18}) and putting, 
\begin{eqnarray}
H(x,y,-y)\equiv \Phi(x,y,-y)+M(y,-y),
\label{eqn:6-21}
\end{eqnarray}
we find
\begin{eqnarray}
{}^gH(x,y,-y)&\equiv& {}^g\Phi(x,y,-y)+M(y,-y)
=g(x,y)H(x,y,-y)g^{\dag}(x,-y).
\label{eqn:6-22}
\end{eqnarray}
The gauge-invariant Dirac action
$S_{{\mbf D}}$
is then given by
\begin{eqnarray}
S_{{\mbf D}}
&=&
\sum_{y=\pm}\int\!d^4x
\big[{\bar\psi}(x,y)i\gamma^\mu(
\partial_\mu+A_\mu(x,y))\psi(x,y)\nonumber\\[2mm]
&&\qquad\;\;\;
+{\bar\psi}(x,y)i\gamma^5H(x,y,-y)\psi(x,-y)\big].
\label{eqn:6-23}
\end{eqnarray}
Setting
\begin{eqnarray}
A_\mu^L(x)&=&A_\mu(x,+),\;\;\;
A_\mu^R(x)=A_\mu(x,-),\nonumber\\[2mm]
H(x)&=&H(+,-),\quad H^{\dag}(x)=H(-,+),
\label{eqn:6-24}
\end{eqnarray}
and
performing the phase
rotation
$\psi_L(x)\to e^{i\pi/4}\psi_L(x)$
and
$\psi_R(x)\to e^{-i\pi/4}\psi_R(x)$,
(\ref{eqn:6-23})
can be put into the well-known form
with Yukawa coupling,
\begin{eqnarray}
S_{{\mbf D}}
&=&
\int\!d^4x
\big[{\bar\psi}_L(x)i\gamma^\mu(
\partial_\mu+A_\mu^L(x))\psi_L(x)
+{\bar\psi}_R(x)i\gamma^\mu(
\partial_\mu+A_\mu^R(x))\psi_R(x)\nonumber\\[2mm]
&&\quad\quad\;\;-{\bar\psi}_L(x)H(x)\psi_R(x)
-{\bar\psi}_R(x)H^{\dag}(x)\psi_L(x)\big].
\label{eqn:6-25}
\end{eqnarray}
\\
\indent
The product rule of the field quantities and
the gauge functions
are now obvious.
Only the functions with the same $x$ and the same 
sign of the variable $y$
can be multiplied.
For instance, there appears no product
like
$H(x,y,-y)\psi(x,y)$.
Similarly, 
only the product like $(\partial_y\Phi(x,y,-y))\psi(y)$ appears.
The same rule should be respected in constructing the bosonic
action.
\\
\indent
To determine the bosonic Lagrangian
we define
the {\it generalized gauge field},\cite{8)}
\begin{eqnarray}
{\mbf A}(x,y,-y)&=&
i\gamma^\mu A_\mu(x,y)
+i\gamma^5\Phi(x,y,-y),\nonumber\\[2mm]
\gamma^0{\mbf A}^{\dag}(x,y,-y)\gamma^0
&=&A(x,y)+i\gamma^5\Phi(x,-y,y).
\label{eqn:6-26}
\end{eqnarray}
The corresponding
generalized field strength
is obtained by
\begin{eqnarray}
{\mbf A}(x,y,-y)=
{\mbf d} {\mbf A}(x,y,-y)
+{\mbf A}(x,y,-y)\wedge{\mbf A}(x,y,-y).
\label{eqn:6-27}
\end{eqnarray}
Here, the generalized
exterior derivative
is defined as
\begin{eqnarray}
{\mbf d}{\mbf A}(x,y,-y)&=&
\frac 12i\gamma^\mu\wedge i\gamma^\nu(\partial_\mu A_\nu(x,y)
-\partial_\nu A_\mu(x,y))\nonumber\\[2mm]
&&+i\gamma^5\wedge i\gamma^\mu\partial_yA_\mu(x,y)
+i\gamma^\mu\wedge i\gamma^5\partial_\mu\Phi(x,y,-y)\nonumber\\[2mm]
&&+
i\gamma^5\wedge i\gamma^5\partial_y\Phi(x,y,-y),
\label{eqn:6-28}
\end{eqnarray}
with the wedge product of $\gamma$ matrices
including $\gamma^5$
being defined in Ref.~18),
so that $\gamma^5\wedge \gamma^\mu=-
\gamma^\mu\wedge \gamma^5=\gamma^5\gamma^\mu$
and $\gamma^5\wedge \gamma^5=1$.
Although the
derivative
$\partial_yA_\mu(x,y)$
is similar to the derivative
(\ref{eqn:6-20}),
the derivative
$\partial_y\Phi(x,y,-y)$
is different.
To explain it
we assume the Leibniz rule
\begin{eqnarray}
\partial_y(\Phi(x,y,-y)\psi(x,-y))&=&
M(y,-y)\Phi(x,-y,y)\psi(x,y)\nonumber\\[2mm]
&=&
-\Phi(x,y,-y)\partial_y\psi(x,-y)
+(\partial_y\Phi(x,y,-y))\psi(x,y).
\label{eqn:6-29}
\end{eqnarray}
Note the minus sign
in front of the first term
of the last expression.
It is now trivial
to
obtain
\begin{eqnarray}
\partial_y\Phi(x,y,-y)
=M(y,-y)\Phi(x,-y,y)
+\Phi(x,y,-y)M(-y,y)
\label{eqn:6-30}.
\end{eqnarray}
We can thus evaluate the exterior derivative
${\mbf d}{\mbf A}(x,y)$.
\\
\indent
The second term in (\ref{eqn:6-27}) is obtained as
\begin{eqnarray}
{\mbf A}(x,y,-y)\wedge {\mbf A}(x,y,-y)&=&
\frac 12i\gamma^\mu\wedge i\gamma^\nu[A_\mu(x,y),
A_\nu(x,y)]\nonumber\\[2mm]
&&+i\gamma^5\wedge i\gamma^\mu\Phi(x,y,-y)A_\mu(x,-y)
\nonumber\\[2mm]
&&+i\gamma^\mu\wedge i\gamma^5
A_\mu(x,y)\Phi(x,y,-y)\nonumber\\[2mm]
&&+
i\gamma^5\wedge i\gamma^5\Phi(x,y,-y)\Phi(x,-y,y).
\label{eqn:6-31}
\end{eqnarray}
Note a consistent sign change of $y$
in accordance with the product rule
mentioned above.
By adding (\ref{eqn:6-31}) to (\ref{eqn:6-28})
we get
\begin{eqnarray}
{\mbf F}(x,y,-y)&=&
\frac 12i\gamma^\mu\wedge i\gamma^\nu F_{\mu\nu}(x,y)
+i\gamma^5\wedge i\gamma^\mu F_\mu(x,y,-y)
+
i\gamma^5\wedge i\gamma^5 F(x,y,-y),
\label{eqn:6-32}
\end{eqnarray}
where\footnote{As noted in Ref.~8)
the component $F(x,y,-y)$
is not gauge-covariant
unless $M(y,-y)M(-y,y)$ commutes with
$g(x,y)$. In the following
we assume
that the matrix $M(y,-y)M(-y,y)$ 
is proportional to the unit matrix. This
assumption
is also necessary
to prove the gauge-Bianchi identity
for the generalized field strength.}
\begin{eqnarray}
F_{\mu\nu}(x,y)&=&\partial_\mu A_\nu(x,y)
-\partial_\nu A_\mu(x,y)+[A_\mu(x,y),
A_\nu(x,y)],\nonumber\\[2mm]
F_\mu(x,y,-y)&=&
\partial_\mu\Phi(x,y,-y)+A_\mu(x,y)\Phi(x,y,-y)
+A_\mu(x,y)M(y,-y)
\nonumber\\[2mm]
&&-\Phi(x,y,-y)A_\mu(x,-y)-M(y,-y)A_\mu(x,-y)
\equiv D_\mu H(x,y,-y),
\nonumber\\[2mm]
F(x,y,-y)&=&
\Phi(x,y,-y)
+M(y,-y)\Phi(x,-y,y)
+\Phi(x,y,-y)M(-y,y)
\nonumber\\[2mm]
&\equiv& H(x,y,-y)H(x,-y,y)-M(y,-y)M(-y,y).
\label{eqn:6-33}
\end{eqnarray}
\indent
Our final task is to construct the bosonic
Lagrangian through the generalized
field strength ${\mbf F}\,(x,y,-y)$.
A method leading to
(\ref{eqn:6-14})
no longer works because
the expression
\begin{eqnarray}
{S'}_{\mbf B}
=-\frac 18\sum_{y=\pm}\int\!d^4x{\rm Tr}\frac 1{g_y^2}
{\mbf F}^{\,2}(x,y,-y),
\label{eqn:6-34}
\end{eqnarray}
contains fewer parameters than
those required by the renormalizability of the
model.
To avoid\cite{19)} the shortcomings
we introduce the associated
field strength with the same gauge transformation
property as that of ${\mbf F}\,(x,y,-y)$,
\begin{eqnarray}
{\tilde{\mbf F}}\,(x,y,-y)
=
\sum_Ah_A^2\Gamma_A{\mbf F}\,(x,y,-y)
\Gamma^A,
\label{eqn:6-35}
\end{eqnarray}
where the sum
over $A$ runs over $S,V,A,T,P$
corresponding to
$\Gamma^A=1,\gamma^\mu,\gamma^5\gamma^\mu,
\sigma^{\mu\nu},
\gamma^5$,
respectively.
We define
the positive parameters $\alpha^2$
and $\beta^2$ in place of $h_A^2$
to write down the bosonic Lagrangian by
\begin{eqnarray}
S_{\mbf B}
&=&-\frac 18\sum_{y=\pm}\int\!d^4x{\rm Tr}\frac 1{g_y^2}
{\tilde{\mbf F}}\,(x,y,-y){\mbf F}\,(x,y,-y)\nonumber\\[2mm]
&=&
-\frac 14\sum_{y=\pm}\int\!d^4x{\rm tr}\frac 1{g_y^2}
F_{\mu\nu}^{\dag}(x,y)F^{\mu\nu}(x,y)\nonumber\\[2mm]
&&+\frac {\alpha^2}2
\sum_{y=\pm}\int\!d^4x{\rm tr}\frac 1{g_y^2}
F_\mu(x,-y,y)F^\mu(x,y,-y)\nonumber\\[2mm]
&&-\frac {\beta^2}2
\sum_{y=\pm}\int\!d^4x{\rm tr}\frac 1{g_y^2}
F(x,-y,y)F(x,y,-y).
\label{eqn:6-36}
\end{eqnarray}
\indent
It would be instructive to
consider a simple chiral gauge group
$U(1)\times U(1)$,
namely, the
Abelian Higgs model coupled to fermion.
After rescaling of the gauge and the Higgs fields
with $g_+=g$ and $g_-=g'$,
and using the notation of
(\ref{eqn:6-24})
we obtain from (\ref{eqn:6-37}),
\begin{eqnarray}
S_{\mbf B}
&=&\int\!d^4x\big[
-\frac 14F_{\mu\nu}^{L\dag}(x)F^{L,\mu\nu}(x)
-\frac 14F_{\mu\nu}^{R\dag}(x)F^{R,\mu\nu}(x)
\nonumber\\[2mm]
&&+(D_\mu H(x))^{\dag}D^\mu H(x)
-\frac \lambda 4(H^{\dag}(x)H(x)-\frac {v^2}2)^2
\big],
\nonumber\\[2mm]
D_\mu H(x)&=&
(\partial_\mu-igA_\mu^L(x)+ig'A_\mu^R(x))H(x).
\label{eqn:6-37}
\end{eqnarray}
The covariant derivatives for the chiral
spinors
are given by
\begin{eqnarray}
D_\mu^L\psi_L(x)&=&
(\partial_\mu-igA_\mu^L(x))\psi_L(x),
\nonumber\\[2mm]
D_\mu^R\psi_R(x)&=&
(\partial_\mu-ig'A_\mu^R(x))\psi_R(x),
\label{eqn:6-38}
\end{eqnarray}
and the fermion mass is proportional
to the parameter $v$.
The gauge field,
\begin{eqnarray}
B_\mu(x)=\frac 1{\sqrt{g^2+g'{}^2}}
(gA_\mu^L(x)
-g'A_\mu^R(x)),
\label{eqn:6-39}
\end{eqnarray}
becomes massive with the mass squared,
$M^2=v^2(g^2+g'{}^2)$,
while the orthogonal combination,
\begin{eqnarray}
A_\mu(x)=\frac 1{\sqrt{g^2+g'{}^2}}
(g'A_\mu^L(x)
+gA_\mu^R(x)),
\label{eqn:6-40}
\end{eqnarray}
remains massless.
\section{Conclusions}
\indent
As pointed out
by Kadyshevskii\cite{3)},
there are two kinds of
Snyder's quantized space-time
where
the triple commutator
between the operator
coordinates
does not vanish.
In one of them,
which was originally proposed by Snyder\cite{2)},
the spatial coordinates have a discrete 
spectrum of eigenvalues of the form,
$na$, where $n$ is an integer,
while the time coordinate has a continuous
spectrum.
The other is opposite:
the spectrum of the time coordinate 
is discrete, while
that of the spatial coordinates is continuous.
The two possible cases
may correspond to the existence
of the $\pm$
spaces introduced in \S4.
As emphasized there,
the $\pm$
spaces are disjoint not connected by
any Lorentz transformation.
The orthogonal
spaces similar
in nature
is utilized in thermofield dynamics
which
introduces non-tilde and tilde states
to express the ensemble averages
in terms of pure states.
\\
\indent
In the DFR algebra,
the triple commutator
between the operator
coordinates
is assumed to vanish,
keeping the canonical commutator
$[{\hat p}^\mu,{\hat x}^\nu]=ig^{\mu\nu}$
intact.
This assumption
allows us to
introduce the non-commutative 
fields
which are subject to
the Lorentz-invariant
non-local interactions.
Although the non-commutative fields have (4+6)-dimensional
coordinates,
the compactification
of the extra 6-dimensions is not needed,
because the fields do not propagate into the 
extra dimensions.
We modified CCZ covariant moment formula
in a nontrivial way
in \S4.
We also showed that
it is possible to
generalize Connes' two-sheeted structure,
$M_4\times Z_2$,
to the non-commutative space-time.
Connes' interpretation that
the Higgs field can be regarded as a kind
of the gauge fields
in the discrete direction
was explicitly displayed
in \S6 by introducing a set of consistent rules
borrowed from the
differential geometry\cite{8)} on $M_4\times Z_2$.
In spite of the fact that
we only rederived the known
QFT models,
we recall that the discrete space-time,
$M_4\times Z_2$, 
proposed by
Connes\cite{7)} in relation
with the standard model
fits nicely to QFT without introducing
extra physical degrees of freedom.
Our introduction of the two-sheeted structure
into the Lorentz-invariant non-commutative
space-time
will permit us to
construct a non-commutative
standard model
with the non-commutative Higgs fields
being interpreted as a kind of the gauge
fields in the discrete direction.
In this sense Connes' theory\cite{7)}
of the standard model
is related to the Lorentz-invariant NCGT.
\\
\indent
From our lesson in \S6 it is apparent that
the set of the monogenic and the dichotomic
fields with appropriate internal
quantum numbers
can be regarded as the set of
the fields considered by Connes\cite{7)}
in his reformulation of the standard model
on $M_4\times Z_2$.\footnote{Or the set
should be identical to those
encoded in Sogami's generalized covariant
derivative.\cite{15)}}
In this respect it should be remembered
that there exist only two non-Abelian charges,
flavor and color, in the
standard model.
This simple fact is mathematically
interpreted that the chiral matter fields
can be regarded as a bi-module
over the flavor-color algebra.\cite{7)}
In the non-commutative regime
this concept would be generalized
to the non-commutative bi-module as 
explained in Ref.~6).
Realistic model construction
along this line of thought
will be a theme in future works.
\appendix
\section{Lorentz Generators of Snyder's and The DFR Algebras}
%
\indent
The generators of the Lorentz group for the Snyder's algebra
were given by (3$\cdot$11). Using the operator
coordinates (3$\cdot$10) they have\cite{2)} the usual expression
in terms of coordinates and momenta,
\begin{eqnarray*}
{\hat M}^{\mu\nu}={\hat x}^\mu{\hat p}^\nu
-{\hat x}^\nu{\hat p}^\mu.
\end{eqnarray*}
\\
\indent
This is no longer the case for the DFR algebra.
In fact, the orbital angular momentum,
\begin{eqnarray*}
{\hat L}^{\mu\nu}={\hat x}^\mu{\hat p}^\nu
-{\hat x}^\nu{\hat p}^\mu,
\end{eqnarray*}
commutes with the operator ${\hat\theta}^{\mu\nu}$
because of (3$\cdot$3) and (3$\cdot$6).
The tensor nature (3$\cdot$2) of the operator ${\hat\theta}^{\mu\nu}$,
therefore, comes from another source.
In this connection we recall that the generators of the Lorentz group
consist of two terms, the orbital angular momentum and the
spin angular momentum.
Thus we may write
\begin{eqnarray*}
{\hat M}^{\mu\nu}&=&{\hat L}^{\mu\nu}+{\hat S}^{\mu\nu},\\[2mm]
{\hat L}^{\mu\nu}&=&{\hat x}^\mu {\hat p}^\nu-{\hat x}^\nu {\hat p}^\mu,
\end{eqnarray*}
where the spin part ${\hat S}^{\mu\nu}$ is chosen such that
the extra terms in the commutator,
\begin{eqnarray*}
[{\hat L}^{\mu\nu},{\hat x}^\rho]&=&i(g^{\nu\rho}{\hat x}^\mu
-g^{\mu\rho}{\hat x}^\nu)
+[{\hat x}^\mu,{\hat x}^\rho]{\hat p}^\nu
-[{\hat x}^\nu,{\hat x}^\rho]{\hat p}^\mu,
\end{eqnarray*}
are precisely cancelled to reproduce the correct commutation relations, 
the second equation of (1$\cdot$4),
\begin{eqnarray*}
[{\hat S}^{\mu\nu},{\hat x}^\rho]&=&-[{\hat x}^\mu,{\hat x}^\rho]
{\hat p}^\nu
+[{\hat x}^\nu,{\hat x}^\rho]{\hat p}^\mu.
\end{eqnarray*}
Similarly from the commutator $[{\hat L}^{\mu\nu},{\hat p}^\rho]$
we should have
\begin{eqnarray*}
[{\hat S}^{\mu\nu},{\hat p}^\rho]&=&0.
\end{eqnarray*}
The ordinary spin angular momentum in ${\hat S}^{\mu\nu}$
commutes with the operator coordinates.
It is then easy to prove (3$\cdot$2) by computing
the commutator $[{\hat M}^{\mu\nu},{\hat\theta}^{\rho\sigma}]
=[{\hat S}^{\mu\nu},{\hat\theta}^{\rho\sigma}]$.
\vspace{5mm}

\end{document}